\documentclass[prl,twocolumn,showpacs,superscriptaddress]{revtex4-1}
\usepackage{hyperref}
\usepackage{amsfonts,amsthm}
\usepackage{epsf}
\usepackage{t1enc}
\usepackage[latin2]{inputenc}
\usepackage{graphicx}
\usepackage{dcolumn}
\usepackage{bm}

\usepackage{amssymb,eso-pic}
\usepackage{latexsym}
\usepackage{tabularx}
\usepackage{amsxtra} 
\usepackage{amsmath,accents}
\usepackage{dcolumn}

\usepackage{ifpdf,epsfig,array,amsmath,amssymb}

\usepackage{psfrag} 

\psfrag{x}{\footnotesize $x$}
\psfrag{y}{\footnotesize $y$}
\psfrag{z}{\footnotesize $z$} 

\psfrag{x=A}{\tiny $(A,0,0)$}
\psfrag{x=-A}{\tiny $(-A,0,0)$}
\psfrag{y=A}{\tiny $(0,A,0)$}
\psfrag{y=-A}{\tiny $(0,-A,0)$}
\psfrag{z=A}{\tiny $(0,0,A)$}
\psfrag{z=-A}{\tiny $(0,0,-A)$}

\psfrag{s1}{\footnotesize $\boldsymbol{\vec{s}}{}^{\hskip.2mm{}^{{}_{\,[1]}}}$}
\psfrag{v1}{\footnotesize $\boldsymbol{\vec{v}}{}^{\hskip.2mm{}^{{}_{[1]}}}$}
\psfrag{d1}{\footnotesize $\boldsymbol{\vec{d}}{}^{\hskip.2mm{}^{{}_{\,[1]}}}$}
\psfrag{s2}{\footnotesize $\boldsymbol{\vec{s}}{}^{\hskip.2mm{}^{{}_{\,[2]}}}$}
\psfrag{v2}{\footnotesize $\boldsymbol{\vec{v}}{}^{\hskip.2mm{}^{{}_{[2]}}}$}
\psfrag{d2}{\footnotesize $\boldsymbol{\vec{d}}{}^{\hskip.2mm{}^{{}_{\,[2]}}}$}

\usepackage[normalem]{ulem}
\usepackage{color}
\definecolor{blue}{rgb}{0,0,1}
\definecolor{red}{rgb}{1,0,0}

\definecolor{DGREEN}{rgb}{0,0.7,0.3}
\definecolor{grey1}{rgb}{0.52, 0.52, 0.51}

\newcommand{\instar}[1]{\accentset{\smash{\raisebox{-0.12ex}{$\scriptstyle\star$}}}{#1}\rule{0pt}{2.3ex}}

\newcommand{\interior}[1]{\accentset{\smash{\raisebox{-0.12ex}{$\scriptstyle\circ$}}}{#1}\rule{0pt}{2.3ex}}

\DeclareFontFamily{OT1}{rsfs}{} \DeclareFontShape{OT1}{rsfs}{m}{n}{
<-7> rsfs5 <7-10> rsfs7 <10-> rsfs10}{}
\DeclareMathAlphabet{\mycal}{OT1}{rsfs}{m}{n}

\def\sc{{\hskip 3.5pt {{}^{{}^{{}_{{}_{\bowtie}}}}} \kern -8.pt{}}}  
\def\SC{{\hskip 3.5pt {{}^{{}^{{}^{{}_{{}_{\bowtie}}}}}} \kern -10.5pt{}}}

\DeclareMathAlphabet{\mathpzc}{OT1}{pzc}{m}{it}

\begin{document}
	
	\title{
 A simple method of constructing binary black hole initial data
		 }

\author{Istv\'{a}n R\'{a}cz}

\affiliation{Max Planck Institute for Gravitational Physics, Albert Einstein Institute, Golm, Germany}

\affiliation{ Wigner Research Center for Physics, Budapest, Hungary}

\date{\today}

\begin{abstract}{\footnotesize
By applying a parabolic-hyperbolic formulation of the constraints and superposing Kerr-Schild black holes, a simple method is introduced to initialize time evolution of binary systems. 
As the input parameters are essentially the same as those used in the post-Newtonian (PN) setup the proposed method interrelates various physical expressions applied in PN and in fully relativistic formulations. The global ADM charges are also determined by the input parameters, and no use of boundary conditions in the strong field regime is made. 
}  
\end{abstract} 
\maketitle


\emph{Introduction.---}Inspiral and merger of binary black holes is of distinguished importance for the emerging field of gravitational wave astronomy. The involved non-linearities necessitate the use of accurate numerical approaches in determining the emitted waveforms. Precision of these simulations, along with their initializations, is of critical importance in enhancing the detection of gravitational wave signals and in deciphering physical properties of their sources.

\smallskip

Initialization is done by solving the Hamiltonian and momentum constraints for a Riemannian metric $h_{ij}$ and a symmetric tensor field $K_{ij}$ both defined on a three-dimensional manifold $\Sigma$. In the vacuum case these constraints read as (see, e.g.~\cite{choquet})
\begin{align} 
{}^{{}^{(3)}}\hskip-1mm R + \left({K^{j}}_{j}\right)^2 - K_{ij} K^{ij}=0 \label{expl_eh}\\
D_j {K^{j}}_{i} - D_i {K^{j}}_{j}=0\,,\label{expl_em}
\end{align}
where ${}^{{}^{(3)}}\hskip-1mm R$ and $D_i$ denote the scalar curvature and the covariant derivative operator associated with $h_{ij}$, respectively. 

\smallskip

A standard approach of solving the constraints is the conformal method. It is based on pioneering inventions made by Lichnerowicz and York \cite{lich,york0}. By replacing the physical metric $h_{ij}$ and the trace free part of $K_{ij}$ with the conformally rescaled fields $\varphi^{4}\,{\widetilde h}_{ij}$ and $\varphi^{-2}\,{\widetilde K}_{ij}$, where ${\widetilde h}_{ij}$ and ${\widetilde K}_{ij}$ are some auxiliary metric and trace free tensor fields, respectively, they could recast the Hamiltonian and momentum constraints as a semilinear elliptic system for the conformal factor $\varphi$ and for a vector potential contributing to the longitudinal part of ${\widetilde K}_{ij}$ \cite{york0,choquet,bs}.  

\smallskip

Our primary aim is to introduce a new method to initializing binary black hole systems by combining advantages of the parabolic-hyperbolic formulation of the constraint equations  \cite{racz_constraints}, and that of superposed Kerr-Schild black holes \cite{i_jeff}. A number of desirable features come with this proposal. For instance, as we do not apply conformal rescalings our variables retain the physically distinguished nature of $h_{ij}$ and $K_{ij}$. 
The use of superposed Kerr-Schild metric requires input parameters such as the rest masses, the sizes and orientations of the displacements, velocities and spins of the involved black holes. As these are essentially the parameters used also in PN description of binaries
interesting interrelations of PN and fully relativistic setups
are provided by the new proposal. In particular, physically relevant expressions of PN may be used to control the orbital properties of the investigated binaries. Remarkably, each of the global ADM charges can also be given in terms of the input parameters \cite{racz_BH_data}. Therefore, it is possible to fix the ADM mass, centre of mass, linear and angular momenta of the binary system in advance of solving the constraints.  

\smallskip

As expected quasi-local quantities, such as  quasi-local masses and spins, or even more complex quantities such as the binding energy can only be determined after the constraints are solved. Specifically,  
in determining the binding energy the tidal deformations of the black holes and the ADM energy have to be known. 
Notable, besides fixing the ADM energy by the input parameters, our new proposal also avoids the use of preconceptions on tidal deformations, which are inevitably involved in other constructions using excision \cite{excision, bonning, bs, harald2, harald}. As recently exactly these assumptions were identified as potential sources of junk radiation  \cite{chu}, it is more than desirable that in our proposal data for the constrained variables has to be fixed only on a topological two-sphere located in the asymptotic region. Their values in the strong field regime---the physically adequate form and measure of tidal deformations depend on them---are then yielded by integrating the evolutionary form of the constraints.  

\smallskip

Notable, the use of the superposed Kerr-Schild metric in our construction is much more intrinsic than in other currently applied methods \cite{bonning, harald2, lovelace, harald}. While in our proposal the auxiliary metric (\ref{eq:bks4}) is used to fix all the freely specifiable variables, in \cite{bonning, harald2, lovelace, harald} even the Kerr-Schild contributions to the conformally rescaled metric and to the mean curvature had to be suppressed in the asymptotic region in order to guarantee physically desirable fall off behavior for the initial data yielded \cite{lovelace}.  


\smallskip

\emph{The parabolic-hyperbolic system.---}For simplicity, assume that $\Sigma$ is smoothly foliated by a one-parameter family $\mycal{S}_\rho$ of two-surfaces that are the $\rho=const$ level surfaces of some smooth function $\rho: \Sigma \rightarrow \mathbb{R}$, 
i.e.~$\Sigma \approx \mathbb{R}\times \mycal{S}$. 


Choose $\rho^i$ to be a vector field on $\Sigma$ such that $\rho^i \partial_i \rho=1$. This vector field decomposes as 
$\rho^i=\widehat{N}\,\widehat n^i+{\widehat N}{}^i$\,,
where $\widehat n_i= \widehat N \partial_i \rho$ is the unit normal to the level surfaces $\mycal{S}_\rho$, $\widehat N^i=\widehat \gamma{}^i{}_j\,\rho^j$ and  $\widehat \gamma{}^i{}_j=\delta{}^i{}_j-\widehat n{}^i\widehat n_j$. 
The metric $h_{ij}$ and the symmetric tensor field $K_{ij}$ can then be given as 
$h_{ij}=\widehat \gamma_{ij}+\widehat  n_i \widehat n_j$ 
and $K_{ij}= \boldsymbol\kappa \,\widehat n_i \widehat n_j  + \left[\widehat n_i \,{\rm\bf k}{}_j  
+ \widehat n_j\,{\rm\bf k}{}_i\right]  + {\rm\bf K}_{ij}$\,,
with $\widehat \gamma_{ij}=\widehat \gamma{}^e{}_i\widehat \gamma{}^f{}_jh_{ef}$, $\boldsymbol\kappa= \widehat n^k\widehat  n^l\,K_{kl}$, ${\rm\bf k}{}_{i} = {\widehat \gamma}^{k}{}_{i} \,\widehat  n^l\, K_{kl}$
and ${\rm\bf K}_{ij} = {\widehat \gamma}^{k}{}_{i} {\widehat \gamma}^{l}{}_{j}\,K_{kl}$. 
In recasting (\ref{expl_eh}) and (\ref{expl_em}) the trace ${\rm\bf K}^l{}_{l}=\widehat\gamma^{kl}\,{\rm\bf K}_{kl}$ and trace free part $\interior{\rm\bf K}_{ij}={\rm\bf K}_{ij}-\tfrac12\,\widehat \gamma_{ij}\,{\rm\bf K}^l{}_{l}$ of ${\rm\bf K}_{ij}$  
will also be applied. 

\vskip0.02cm
In terms of the variables $\widehat N, \widehat N^i, \widehat \gamma_{ij}; \boldsymbol\kappa, {\rm\bf k}{}_{i},$ $ {\rm\bf K}^l{}_{l}$ and $\interior{\rm\bf K}_{ij}$ the parabolic-hyperbolic form of the Hamiltonian and momentum  constraints 
 can be given for $\widehat N,{\rm\bf k}{}_{i}$ and ${\rm\bf K}^l{}_{l}$ as \cite{racz_constraints} 
\begin{widetext}\vskip-0.3cm
\begin{align}
{}& \instar{K}\,[\,(\partial_{\rho} \widehat N) - \widehat N{}^l(\widehat D_l\widehat N) \,] = \widehat N^{2} (\widehat D^l \widehat D_l \widehat N) + \mathcal{A}\,\widehat N + \mathcal{B}\,\widehat N{}^{3} \,, \label{bern_pde} \\ 
{}& \mycal{L}_{\widehat n} {\rm\bf k}{}_{i} - \tfrac12\,\widehat D_i ({\rm\bf K}^l{}_{l}) - \widehat D_i\boldsymbol\kappa + \widehat D^l \interior{\rm\bf K}{}_{li} + \widehat N \instar{K}\,{\rm\bf k}{}_{i}  + [\,\boldsymbol\kappa-\tfrac12\, ({\rm\bf K}^l{}_{l})\,]\,\dot{\widehat n}{}_i - \dot{\widehat n}{}^l\,\interior{\rm\bf K}_{li} = 0 \label{par_const_n} \\
{}& \mycal{L}_{\widehat n}({\rm\bf K}^l{}_{l}) - \widehat D^l {\rm\bf k}_{l} - \widehat N \instar{K}\,[\,\boldsymbol\kappa-\tfrac12\, ({\rm\bf K}^l{}_{l})\,]  + \widehat N\,\interior{\rm\bf K}{}_{kl}\instar{K}{}^{kl}  + 2\,\dot{\widehat n}{}^l\, {\rm\bf k}_{l}  = 0\,, \label{ort_const_n}
\end{align}
\end{widetext}
where $\widehat D_i$ denotes the covariant derivative operator associated with $\widehat \gamma_{ij}$, \,$\dot{\widehat n}{}_k={\widehat n}{}^lD_l{\widehat n}{}_k=-{\widehat D}_k(\ln{\widehat N})$, 
$\mathcal{A} =(\partial_{\rho} \instar{K}) - \widehat N{}^l (\widehat D_l \instar{K}) + \tfrac{1}{2}[\,\instar{K}^2 + \instar{K}{}_{kl} \instar{K}{}^{kl}\,]$, 
$\mathcal{B} = - \tfrac12\,\bigl[\widehat R + 2\,\boldsymbol\kappa\,({\rm\bf K}^l{}_{l})+\tfrac12\,({\rm\bf K}^l{}_{l})^2 -2\,{\rm\bf k}{}^{l}{\rm\bf k}{}_{l}  - \interior{\rm\bf K}{}_{kl}\,\interior{\rm\bf K}{}^{kl}\,\bigr]$,
$\instar{K}_{ij}=\tfrac12\mycal{L}_{\rho} {\widehat \gamma}_{ij} -\widehat  D_{(i}\widehat N_{j)}$ 
and 
\begin{equation}
\instar{K}  =\tfrac12\,{\widehat \gamma}^{ij}\mycal{L}_{\rho} {\widehat \gamma}_{ij} -  \widehat D_j\widehat N^j\,.\label{trhatext}
\end{equation}

As no restriction applies to 
$\widehat N^i,\widehat \gamma_{ij}, \boldsymbol\kappa$ and $\interior{\rm\bf K}_{ij}$ they are freely specifiable throughout $\Sigma$.
It was also shown in \cite{racz_constraints} that (\ref{bern_pde}) is uniformly parabolic in those subregions of $\Sigma$ where $\instar{K}$ is either positive or negative. Note also that $\instar{K}$ depends exclusively on the freely specifiable fields $\widehat \gamma_{ij}$ and $\widehat N^i$ \cite{racz_constraints}.

It was also shown in \cite{racz_constraints} that 
if suitable initial values for the constrained fields 
are given, on some level surface $\mycal{S}_0$ in $\Sigma$, then, in the domain of dependence of $\mycal{S}_0$, unique solution exists to the evolutionary system (\ref{bern_pde})--(\ref{ort_const_n}) such that the fields $h_{ij}$ and $K_{ij}$ that can be reconstructed there, from the free data and constrained variables do satisfy (\ref{expl_eh}) and (\ref{expl_em}).



\medskip

\emph{Kerr black holes in Kerr-Schild form.---}A Lorentzian metric $g_{\alpha\beta}$ 
is of Kerr-Schild type if it is of the form 
\begin{equation}\label{eq:ksm}
         g_{\alpha\beta}=\eta_{\alpha\beta}+2 H \ell_{\alpha} \ell_{\beta}\,, 
\end{equation}
or equivalently, in inertial coordinates $(t,x^i)$ adapted to the background Minkowski metric $\eta_{\alpha\beta}$, it can be given as
\begin{align}
g_{\alpha\beta}\,dx^{\alpha} dx^{\beta}= {}&  (-1+2H{\ell_0}^2)\,dt^2 + 4H \ell_0\ell_i \,dt dx^i \nonumber \\ {}&  + (\delta_{ij}+ 2 H \ell_i \ell_j)\,dx^i dx^j\,, 
\end{align}
where $H$, apart from singularities, is a smooth function on $\mathbb{R}^4$ and
$\ell_{\alpha}$ is null with respect to both $g_{\alpha\beta}$ and $\eta_{\alpha\beta}$. In particular, for $\ell^{\alpha}=g^{\alpha\beta}\ell_{\beta} = \eta^{\alpha\beta}\ell_{\beta}$ the relations $g^{\alpha\beta}\ell_{\alpha}\ell_{\beta} = \eta^{\alpha\beta}\ell_{\alpha}\ell_{\beta}=-(\ell_0)^2+\ell^i\ell_i=0$ and $\ell^{\beta} \partial_{\beta} \, \ell^{\alpha}=0$ hold.  

\smallskip


The  Kerr black hole \cite{ks1} is of Kerr-Schild form with
\begin{equation}\label{H-ell-kerr}
H=\frac{r^3M}{r^4+{a^2z^2}} \,\,\, {\rm and} \,\,\, \ell_{\alpha}=\left(1, \frac{r\,x+a\,y}{r^2+a^2},
\frac{r\,y-a\,x}{r^2+a^2},\frac{z}{r}  \right)\,,
\end{equation}
where the Boyer-Lindquist radial coordinate $r$ is related to 
the spatial part of the inertial coordinates $x^i =(x,y,z)$ as 
\begin{equation}\label{imp-r-def}
r^4-(x^2+y^2+z^2-a^2)\,r^2-a^2\,z^2=0\,.
\end{equation}
The ADM mass, centre of mass, linear and angular momenta of asymptotically flat solutions can be determined by applying the asymptotic expansions. In particular, for the Kerr-Schild black hole, given by (\ref{H-ell-kerr}) and (\ref{imp-r-def}), the ADM mass is $M$, the centre of mass is represented by the origin of the background Euclidean space, the linear momentum vanishes (either of the latter two properties means that the black hole is in rest with respect to the background reference frame), while the ADM angular momentum is $\vec{J}=a M \vec{e}_z$, where the unit vector $\vec{e}_z$ points to the positive $z$ direction. 


\medskip

\emph{Generic Kerr-Schild black holes.---}The most important advances in using Kerr-Schild metrics come with their form-invariance under Lorentz transformations. Accordingly, if a Lorentz transformation $x'{}^{\alpha}=\Lambda^{\alpha}{}_{\beta}\,x^{\beta}$ is performed
the metric retains its distinguished form $g'_{\alpha\beta}=\eta_{\alpha\beta}+2 H' \ell'_{\alpha} \ell'_{\beta}$, where $H'=H'(x'{}^{\alpha})$ and $\ell'_{\beta}=\ell'_{\beta}(x'{}^{\varepsilon})$ are given as
\begin{equation}
	H' =  H\left([\Lambda^{\alpha}{}_{\beta}]^{-1}x'{}^{\beta}\right),\  \ell'_{\beta} = \Lambda^{\alpha}{}_{\beta}\, \ell_{\alpha}\left([\Lambda^\varepsilon{}_\varphi]^{-1}x'{}^\varphi\right)\,.
\end{equation}

\smallskip

Since boosts and rotations are special Lorentz transformations it is straightforward to construct moving and rotating black holes with preferably oriented speed and spin by performing suitable sequence of boosts and rotations starting with a Kerr black hole. 

As a simple example consider a Kerr black hole that is in rest with respect to some reference system $x'{}^{\alpha}$. Then, $H(x^{\alpha})$ and $\ell_{\alpha}(x^{\varepsilon})$, relevant for a black hole that is displaced by distance $d$ in the positive $y$ direction and moving with velocity $0< v < 1$ in the positive $x$ direction of a reference system $x{}^{\alpha}$, are obtained by substituting $x'=\gamma\, x - \gamma v\, t$, $y'= y - d$ and $z'=z$ into
\begin{align}
H= {}& \frac{r'{}^3M}{r'{}^4+{a^2 z'{}^2}}  \quad {\rm and}\\ \ell_{\beta}={}& \left(\gamma\,\ell'_0 - \gamma v \, \ell'_1, \gamma \,\ell'_1 - \gamma v \,\ell'_0, \ell'_2,\ell'_3\right) \,,
\end{align}
where $\gamma=1/\sqrt{1-v^2}$, while $\ell'_{\beta}$ and $r'{}$ are determined by the primed variant of (\ref{H-ell-kerr}) and (\ref{imp-r-def}), respectively.

Asymptotic expansions, in accordance with the transformations preformed, verify that for the considered displaced, boosted and spinning black holes the ADM mass, centre of mass, linear and angular momenta can be given as $\gamma\,M$, $\vec{d}$, $\gamma\,M\,\vec{v}$ and $\gamma\,M\{\vec{d} \times\vec{v} + a\,\vec{e}_z \}$, respectively, where $\vec{d}=d\,\vec{e}_y$, $\vec{v}=v\,\vec{e}_x$, and the unit vectors  $\vec{e}_x$ and $\vec{e}_y$ are aligned to the positive $x$ and $y$ directions, respectively.
 

\medskip

\emph{Superposed Kerr-Schild black holes.---}The metric of binaries, composed by two moving and spinning black holes, will be approximated by
\begin{equation}\label{eq:bks4}
g_{\alpha\beta}= \eta_{\alpha\beta} + 2 H {}^{[1]}\ell_{\alpha} {}^{[1]}\ell_{\beta}{}^{[1]} + 2 H {}^{[2]}\ell_{\alpha} {}^{[2]}\ell_{\beta}{}^{[2]}\,,
\end{equation}
where $ H {}^{[n]}$ and $\ell_\alpha {}^{[n]}$ correspond to the Kerr-Schild data for individual black holes. 

\smallskip

If (\ref{eq:bks4}) solved Einstein's equations then the inertial three-metric 
\begin{equation}\label{eq:bks3}
h_{ij}= \delta_{ij} + 2 H {}^{[1]}\ell_i {}^{[1]}\ell_j{}^{[1]} +2 H {}^{[2]}\ell_i {}^{[2]}\ell_j{}^{[2]}\,,
\end{equation}
and the extrinsic curvature $K_{ij}$ that could be deduced from (\ref{eq:bks4}) would satisfy the constraint equations on $t=const$ hypersurfaces and, in turn, the corresponding fields $\widehat N,\widehat N^i,\widehat \gamma_{ij},\boldsymbol\kappa, {\rm\bf K}^l{}_{l},{\rm\bf k}{}_{i}$ and $\interior{\rm\bf K}_{ij}$ would also satisfy (\ref{bern_pde})--(\ref{ort_const_n}). 

\smallskip

Although the auxiliary metric (\ref{eq:bks4}) does not solve Einstein's equations it is known to be a good approximation close to the individual black holes 
\cite{i_jeff}. Direct calculation also verifies that in the asymptotic region the Einstein tensor falls off as $\mathcal{O}(|\vec{x}|^{-4})$, where $|\vec{x}|=\sqrt{x^2+y^2+z^2}$. Whence, it is more than tempting to choose, on $t=const$ hypersurfaces, the freely specifiable fields $\widehat N^i,\widehat \gamma_{ij},\boldsymbol\kappa$ and $\interior{\rm\bf K}_{ij}$ as if (\ref{eq:bks4}) solved the Einstein equations. Recall that equations (\ref{bern_pde})--(\ref{ort_const_n}) require initialization of the constrained fields $\widehat N$,  ${\rm\bf k}{}_{i}$ and ${\rm\bf K}^l{}_{l}$ on one of the level surfaces, say on $\mycal{S}_0$, which is also done by applying (\ref{eq:bks4}). As seen below, these choices will be approved by significant paybacks. 

\smallskip

So far the free data has been chosen by using the auxiliary metric (\ref{eq:bks4}). Note, however, that as the metric (\ref{eq:bks4}) does not solve Einstein's equations, the true solutions $\widehat N$,  ${\rm\bf k}{}_{i}$ and ${\rm\bf K}^l{}_{l}$ to the evolutionary system (\ref{bern_pde})--(\ref{ort_const_n}) will always differ, in the interior of the domain of dependence of $\mycal{S}_0$, from those fields that could be deduced from (\ref{eq:bks4}). 


\smallskip

\emph{The boundary-initial value problem.---}Hitherto the level surfaces $\mycal{S}_\rho$ have tacitly been assumed to be compact without boundary. However, in most of the numerical approaches the initial data surface $\Sigma$ is chosen to be a sufficiently large but bounded subset of $\mathbb{R}^3$. 
In adopting such a scheme here the product structure $\Sigma \approx \mathbb{R}\times \mycal{S}$ will be guaranteed by applying leaves $\mycal{S}_\rho$ that are diffeomorphic to a closed disk in $\mathbb{R}^2$.

\begin{figure}[htb]
	\begin{center}
		\includegraphics[width=7.3cm]{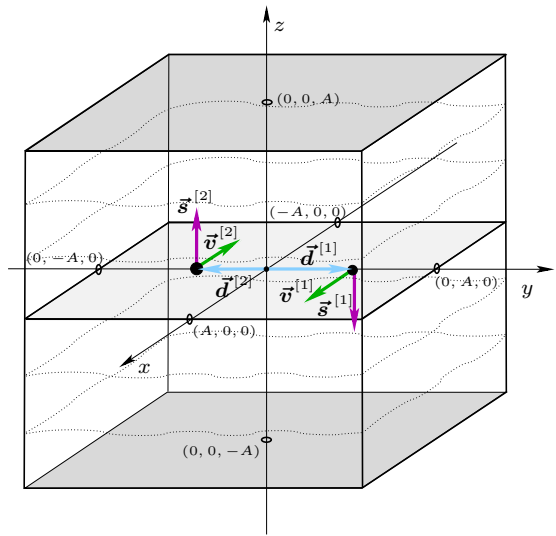}
	\end{center}
	\vskip-.5cm\caption{\footnotesize{(color online). The initial data surface $\Sigma$ is chosen to be the cube centered at the origin in $\mathbb{R}^3$ with edges $2A$. It will be argued below that initial data 
			can be specified on the horizontal squares, with $z=\pm A$, bounding the cube from above and below, whereas boundary values can be given on the complementary part of the boundary comprised by four vertical squares.}}
	\label{kocka}
\end{figure}

Here we choose $\Sigma$ to be the cube (see Fig.\,\ref{kocka}) centered at the origin in $\mathbb{R}^3$ with edges $2A$, which, for sufficiently large value of $A$ contains the binary system with a reasonable size of margin.  
The price for doing this is that the parabolic-hyperbolic system (\ref{bern_pde})--(\ref{ort_const_n}) has to be solved as an initial-boundary value problem. It is important that if (\ref{bern_pde}) is uniformly parabolic well-posedness of such a problem is guaranteed (see, e.g.~\cite{kreissl}), though, a suitable splitting of the boundary of $\Sigma$ into disjoint subsets on which the initial and boundary values can be specified, respectively, has also to be find. 

\smallskip

Before splitting the boundary of $\Sigma$, consisting of six squares, into suitable parts where initial and boundary values are to be specified recall that (\ref{bern_pde}) is uniformly parabolic only in those subsets of $\Sigma$, where $\instar{K}$ is strictly negative or positive. Indeed, it is the sign of $\instar{K}$ that decides whether the system (\ref{bern_pde})--(\ref{ort_const_n}) evolves in the positive or negative $\rho$-direction. It propagates aligned the vector filed $\rho^i$ for positive $\instar{K}$, while anti-aligned for negative $\instar{K}$.

Restrict now considerations to a binary with speeds, displacements and spins aligned parallel to the $x,y$ and $z$-axis, respectively (as indicated on Fig.\,\ref{kocka}). Apply then a foliation of $\Sigma$ by $z=const$ level surfaces, and determine the function $\instar{K}$ using (\ref{eq:bks4}).
Direct calculation verifies then that $\instar{K}$ can be given as the product of a strictly negative function and the $z$-coordinate. (Note that all the point-like or ring-like singularities are now confined to the $z=0$ plane.) This means that $\instar{K}$ is positive everywhere below the $z=0$ plane while it is negative above that plane. This behavior may also be verified by plotting $\instar{K}=const$ level surfaces as it is done for a specific choice of physical parameters  on Fig.~S.1 of the Supplemented Material \cite{SM1}. 


Then, for a binary black hole system arranged as indicated on Fig.\,\ref{kocka}, the evolutionary equations (\ref{bern_pde})--(\ref{ort_const_n}) are well-posed on the disjoint domains, $\Sigma^+$ and $\Sigma^-$, above and below the $z=0$ plane. 
In particular, they may be solved by propagating initial values specified on the horizontal $z=\pm A$ squares, along the $z$-streamlines, meanwhile the $z=const$ `time' level surfaces approach the orbital plane 
from above and below. The boundary values are to be given on the four vertical sides of the cube (see Fig.\,\ref{kocka}). As the fields $\widehat N$,  ${\rm\bf k}{}_{i}$ and ${\rm\bf K}^l{}_{l}$ are developed on $\Sigma^+$ and $\Sigma^-$ separately the existence of sufficiently smooth unique solutions on the union of the closures of $\Sigma^+$ and $\Sigma^-$, respectively, is of fundamental importance. 


Notably, there is a significant simplification offered by the specific choice we made for restricted class of binary black hole systems, and also by fixing the freely specifiable part of data using (\ref{eq:bks4}). Indeed, it is straightforward to check that for the considered class of binary black hole configurations, the auxiliary metric (14) possesses a $z \rightarrow -z$ reflection symmetry. This, in particular, along with a suitably iterative argument based on results covered by \cite{bartnik,lieberman, weinstein}, can be used to verify both the existence and uniqueness of ``global'' solutions on $\Sigma^+$ and $\Sigma^-$, and also (apart from singularities) their sufficiently smooth matching at the $z=0$ plane. [For more details see the Supplemental Material \cite{SM1}.]


\medskip
 
\emph{Conclusions.---}By applying the parabolic-hyperbolic formulation of constraints and superposed Kerr-Schild black holes a radically new method to construct binary black hole initial data was introduced. 
The main advantages of the proposed new method come with its simplicity, with the intimate interrelation of the applied input parameters---the rest masses, velocities, spins and  displacements of the individual black holes---and those used in the PN formalism. The latter, along with some PN relations, could be used to fix orbital properties of binaries.  In addition,---as shown in \cite{racz_BH_data} (see also the Supplementary Material \cite{SM1})---all of the global ADM charges can be given by linear combination of those for
individual black holes, thereby they are also determined the input parameters. It is also remarkable that instead of involving any sort of preconception on tidal deformations the new proposal determines the adequate contributions.

\smallskip

A major motivation for this paper is to encourage numerical implementations which will be important in extending the  merits of the proposed analytic setup. Numerical simulations will be needed to determine quasilocal quantities or the binding energy, as well as, to explore the functional dependence of these quantities on the input parameters.

\smallskip

Notably, the superposed Kerr-Schild metric (\ref{eq:bks4}) has no gravitational wave content. Accordingly, the radiative degrees of freedom are turned on only via the four constrained variables, $\widehat N,{\rm\bf k}{}_{i}$ and ${\rm\bf K}^l{}_{l}$. This should yield a minimizing of the spurious radiation content, though 
careful numerical investigations of time evolutions will be needed to see if a desired suppressing of junk radiation will indeed occur.

\smallskip

Note, finally, that for definiteness we treated here only the case of binary black holes. Nevertheless, our proposal immediately applies to multiple systems whenever the initial speeds are parallel to the $x-y$-plane and the spins are orthogonal to that plane. As there are no restrictions, besides some obvious ones, on the input parameters this set hosts a great number of multiple black hole configurations of immediate physical interest.


\smallskip

The author is grateful to  Lars Andersson, Harald Pfeiffer, Bob Wald and Jeff Winicour for helpful comments. This work was supported in part by the NKFIH grant K-115434. 
  

\end{document}